\documentclass[12pt]{article}
\usepackage{amsfonts}
\usepackage{amsmath}
\usepackage{amssymb}
\usepackage{amstext}
\usepackage{graphicx,epsfig}
\usepackage{hyperref}
\usepackage[nooneline]{caption2}
\makeatletter
\def\ExtendSymbol#1#2#3#4#5{\ext@arrow 0099{\arrowfill@#1#2#3}{#4}{#5}}
\def\RightExtendSymbol#1#2#3#4#5{\ext@arrow 0359{\arrowfill@#1#2#3}{#4}{#5}}
\def\LeftExtendSymbol#1#2#3#4#5{\ext@arrow 6095{\arrowfill@#1#2#3}{#4}{#5}}

\makeatother

\begin{document}
\baselineskip 20pt

\title{Scalable Generation and Characterization of a Four-Photon Twelve-Qubit
Hyperentangled State}
\author{Kun Du$^a$ and
Cong-Feng Qiao$^{a,b}$\footnote{Corresponding author. Email: qiaocf@gucas.ac.cn}\\[0.5cm]
$^{a}$Department of Physics, Graduate University of
Chinese Academy of Sciences,\\ Beijing 100049, China\\[0.2cm]
$^{b}$Theoretical Physics Center for Science Facilities (TPCSF),\\
CAS, Beijing 100049, China}

\date{}
\maketitle

\begin{abstract}
\noindent
An experimentally feasible scheme for generating a 12-qubit
hyperentangled state via four photons, entangled in polarization,
frequency and spatial mode, is proposed. We study the nature of
quantum non-locality of this hyperentangled state by evaluating its
violation degree to a Bell-type inequality, and find that the result
agrees well with quantum mechanics prediction while extremely
contradicts to the local realism constraint.\\\\
\textbf{Keywords:} hyperentanglement; Bell inequality; non-locality;
Qubit
\end{abstract}

\section{Introduction}

Qubit is the basic information unit of quantum computer. Hence, the
generation of multi-qubit entangled states keeps as an essential
task in quantum information processing and transmission
\cite{quantum-computation}. Quantum entanglement, especially
multipartite entanglement, has many important applications, such as
measurement-based quantum computing \cite{one-way quantum computer},
secure superdense coding \cite{C. Wang et al}, teleportation
\cite{Teleporting}, entanglement purification \cite{Purification
using Spatial Entanglement} and quantum cryptography \cite{D. Bruss
and C. Macchiavello}. However, the experimental realization of
multipartite entanglement is still a significant challenge. People
realized that the manipulation of more than six photons without
quantum storage turns out to be an insurmountable obstacle with
present technology \cite{nature Pan}. However, the
hyperentanglement(HE), which in practice remains in many physical
systems, tends to be an effective and practical way \cite{Kwiat P.
G,Ultrabright source,Polarization-momentum} to manipulate more
qubits effectively, namely making the quanta, e.g. photons, to be
entangled simultaneously in multiple degrees of freedom (DOFs).
Furthermore, HE states are much less affected by decoherence and are
particular important in the realization of even more challenging
objectives in comparison with the normal entangled states.

Recently, two-photon four-qubit states \cite{two-photon
four-qubit,All-Versus-Nothing,All-Versus-Nothing2} and two-photon
six-qubit states \cite{two-photon six-qubit,Hyperentangled Photon
Pairs} were experimentally realized, and found to offer significant
enhancement of the channel capacity in superdense coding
\cite{Schuck-96-2006,Barreiro-4-2008}, to construct efficiently the
fourqubit two-photon cluster states \cite{Vallone-100-2008}, and to
realize successfully the basic quantum computation algorithms in the
one-way model \cite{two-photon four-qubit,Briegel-86-2001}. More
recently, Jian-Wei Pan {\it et al.} generated in experiment the
hyperentangled ten-qubit photonic Schr\"odinger cat state by
exploiting polarization and momentum of five photons \cite{nature
Pan}. In this article, we propose an experimentally practical scheme
for generating a four-photon twelve-qubit hyperentangled state
entangled in three DOFs, i.e. in polarization, frequency and spatial
mode.

Nonlocality is a unique feature of Quantum Mechanics(QM) differing
from the local realism(LR) theory. People found that the HE states
can enhance the violation of LR \cite{Barbieri-97-2006}. We find in
this work that the twelve-qubit hyperentangled state exhibits
extremely large violation of the LR constraint by evaluating a
Bell-type inequality. Furthermore, we show generally that the more
the entangled degrees of freedom are, the stronger the violation of
Bell-type inequality will be

The structure of the paper goes as follows: in section 2, an
experimentally feasible scheme for the production and measurement of
a four-photon twelve-qubit hyperentangled state is proposed. In
section 3, we discuss the quantum characteristics of the
hyperentangled state(mathematic details are presented in Appendix).
A brief summary and concluding remarks are given in section 4.

\section{Preparation scheme for hyperentangled state}

The hyperentangled state to be generated in our scheme takes the
following form up to a normalization constant:
\begin{eqnarray}
|\psi\rangle=(|HHHH\rangle+|VVVV\rangle)\bigotimes (|\omega_{1}
\omega_{2}\omega_{1}\omega_{2}\rangle +
|\omega_{2}\omega_{1}\omega_{2}\omega_{1}\rangle) \bigotimes
(|a'_{1}b'_{1}c'_{1}d'_{1}\rangle+|a'_{2}b'_{2}c'_{2}d'_{2}\rangle)\;
,\label{shi1}
\end{eqnarray}
where $H$ and $V$ denote horizontal and vertical polarization,
$\omega_{1}$ and $\omega_{2}$ signify different frequency and
$a'_{1}$, $b'_{1}$, $c'_{1}$, $d'_{1}$, $a'_{2}$, $b'_{2}$,
$c'_{2}$, $d'_{2}$ label eight different spatial modes. The state
(\ref{shi1}) shows maximal entanglement among all photons' degrees
of freedom in polarization, frequency and spatial modes. This state
is a Schr\"{o}inger cat state, and can also be called as
Greenberger-Horne-Zeilinger state \cite{ghz1990}, which possesses
peculiar interest in quantum information science.

\begin{figure}[h]
\centering
\includegraphics[width=9.5cm]{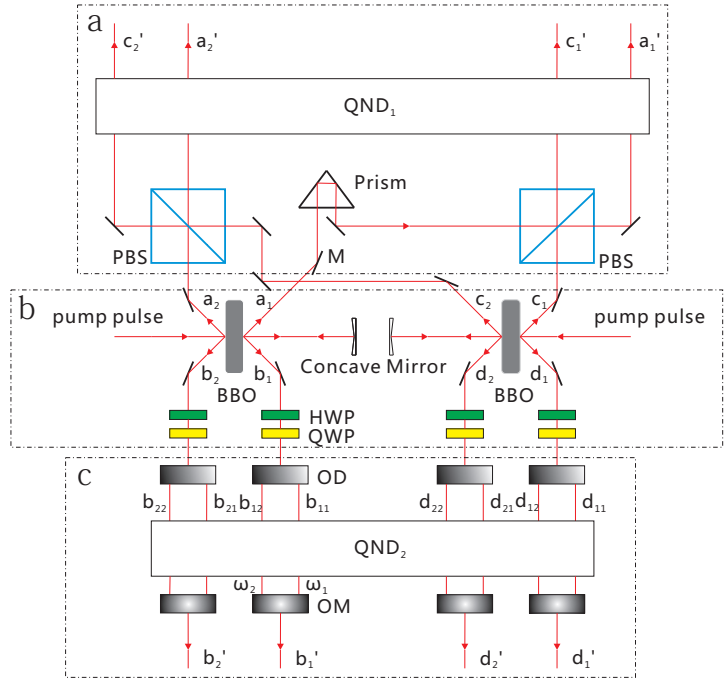}
\caption{Preparation scheme. a) SPDC source. The two sets of crystal
system generate photon pairs entangled in polarization, frequency
and spatial mode respectively. b) Design of constructing the
entanglement of four photons with one another in polarization and
spatial mode. Two PBSs are exploited to implement the selection of
polarization, and a cross-kerr nonlinear medium which can accomplish
the function as quantum nondemolition detector (QND) is used to
choose photons' spatial modes. c) Design of constructing the
entanglement of four photons with one another in frequency and
spatial mode. Analogously, the selection of frequency is achieved by
four sets of Wavelength Division Multiplexer (WDM) which consists of
optical demultiplexer (OD) and optical multiplexer (OM). Moreover,
the complete experimental set still needs one more cross-kerr
nonlinear medium to choose photons' spatial modes the same as
explained in above.} \label{fig1}
\end{figure}

To generate the state, the first step is to produce two pairs of
two-dimensional hyperentangled photons, as shown in Figure
\ref{fig1}a, via two adjacent thin type-II barium borate (BBO)
crystals \cite{BBO}. The photon pairs can be generated by
spontaneous parametric down-conversion (SPDC) \cite{SPDC} and
entangled in polarization and frequency DOFs
\cite{frequency-entangled}. If pump photon downconvert in first
(second) crystal, the frequency of the signal photon will be
$\omega_{1}$ ($\omega_{2}$), and hence the Idler photon takes the
opposite. In case a pump pulse coming from left (right) threads the
first (second) set of crystal, a pair of photons will be produced in
the spatial modes $a_{1}$ and $b_{1}$ ($c_{1}$ and $d_{1}$).
Similarly, a pair of hyperentangled photons can be produced in
spatial modes $a_{2}$ and $b_{2}$ ($c_{2}$ and $d_{2}$)
\cite{Purification using Spatial Entanglement} while the pump pulse
is reflected by the mirror behind the crystal and then threads the
crystal second time. Since one can simply adjust the relative phase
between the first and second possibilities, two pairs of photons in
the following form are readily obtained:
\begin{eqnarray}
(|HH\rangle+|VV\rangle)\bigotimes
(|\omega_{1}\omega_{2}\rangle+|\omega_{2}\omega_{1}\rangle)
\bigotimes (|a_{1}b_{1}\rangle+|a_{2}b_{2}\rangle)\; ,\label{shi2}
\end{eqnarray}
\begin{eqnarray}
(|HH\rangle+|VV\rangle)\bigotimes
(|\omega_{1}\omega_{2}\rangle+|\omega_{2}\omega_{1}\rangle)
\bigotimes (|c_{1}d_{1}\rangle+|c_{2}d_{2}\rangle)\; .\label{shi3}
\end{eqnarray}

\begin{figure}[h]
\centering
\includegraphics[width=8cm]{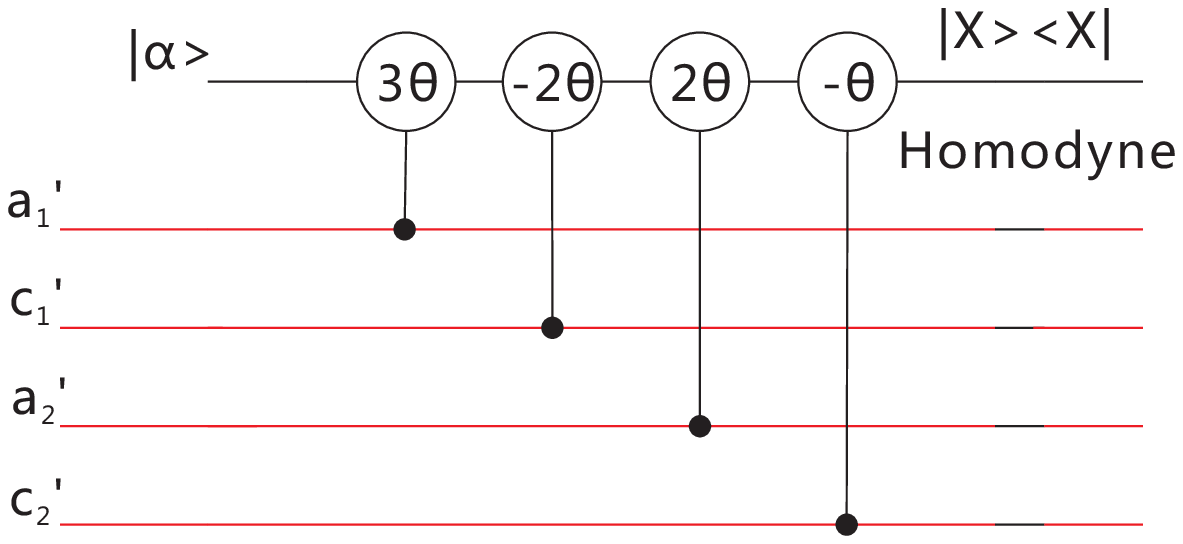}
\caption{$QND_{1}$. The realization of selecting paths of photons a
and c with cross-kerr nonlinear medium.} \label{fig2}
\end{figure}

Next, we employ the two non-entangled photons $a$ and $c$ in above
two pairs to make all photons entangled in polarization and spatial
mode. Obviously, photons $a$ and $c$ are in the state of the form
\begin{eqnarray}
[(|H\rangle+|V\rangle)\bigotimes
(|a_{1}\rangle+|a_{2}\rangle)]\bigotimes
[(|H\rangle+|V\rangle)\bigotimes (|c_{1}\rangle+|c_{2}\rangle)]\;
.\label{shi4}
\end{eqnarray}
As shown in Figure \ref{fig1}b, photons in paths $a_{1}$ and $c_{1}$
($a_{2}$ and $c_{2}$) frontally enter Polarizing Beam Splitters
(PBSs). Since the PBS transmits H and reflects V polarization
photons, we can then come to a conclusion that if the two photons
pass through paths $a'_{1}$ and $c'_{1}$ ($a'_{2}$ and $c'_{2}$)
simultaneously, it implies that both of the photons are H or V
polarized. Afterwards, photons in modes $a'_{1}$, $c'_{1}$, $a'_{2}$
and $c'_{2}$ are led to a cross-kerr nonlinear medium
\cite{cross-kerr,quantum communication} as shown in Figure
\ref{fig2}, which brings forth an adjustable phase shift to the
coherent state $|\alpha\rangle$ through cross-phase modulation
(XPM), e.g. $3\theta$, $-2\theta$, $2\theta$ and $-\theta$ for four
modes, respectively. Then we perform the $X$ homodyne detection on
the coherent state $|\alpha\rangle$ \cite{homodyne}. In case the
measured phase shift is $\theta$, the state will be kept. In the
end, the reserved states of photons $a$ and $c$ can only come out
from $a'_{1}$ and $c'_{1}$ or $a'_{2}$ and $c'_{2}$. That means the
states of photons $a$ and $c$ are in the following entangled form:
\begin{eqnarray}
(|HH\rangle+|VV\rangle)\bigotimes
(|a'_{1}c'_{1}\rangle+|a'_{2}c'_{2}\rangle)\; .\label{shi5}
\end{eqnarray}

\begin{figure}[h]
\centering
\includegraphics[width=12cm]{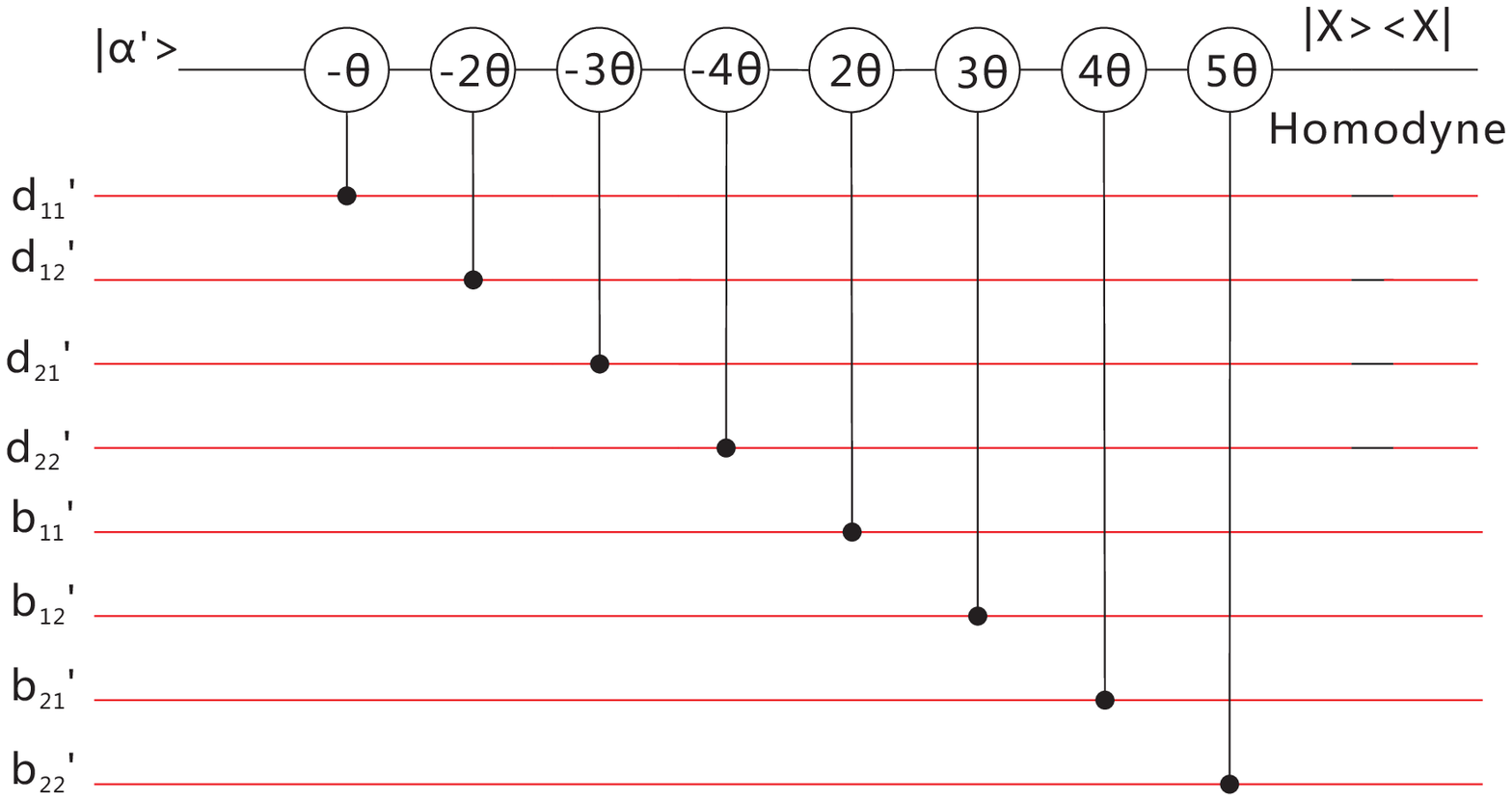}
\caption{$QND_{2}$. The realization of selecting paths of photons b
and d with cross-kerr nonlinear medium.} \label{fig3}
\end{figure}

Similarly, we can use another pair of non-entangled photons $b$ and
$d$ to make all photons entangled in spacial mode and frequency. As
mentioned in above, the initial photons $b$ and $d$ are in the
state:
\begin{eqnarray}
[(|\omega_{1}\rangle+|\omega_{2}\rangle)\bigotimes
(|b_1\rangle+|b_2\rangle)]\bigotimes
[(|\omega_{1}\rangle+|\omega_{2}\rangle)\bigotimes
(|d_1\rangle+|d_2\rangle)]\; .\label{shi6}
\end{eqnarray}
Connecting paths $b_{1}$, $b_{2}$, $d_{1}$ and $d_{2}$ with four
optical demultiplexers (OD) respectively, which are part of
Wavelength Division Multiplex (WDM) \cite{WDM}, we then separate
each mode into two in terms of frequency as shown in Figure
\ref{fig1}c. The modes $b_{11}$, $b_{21}$, $d_{11}$ and $d_{21}$
correspond to frequency $\omega_{1}$, while the modes $b_{12}$,
$b_{22}$, $d_{12}$ and $d_{22}$ correspond to $\omega_{2}$. Taking
the same procedure as for modes a and c, we lead these eight paths
to another cross-kerr nonlinear medium, as shown Figure \ref{fig3},
which induces the coherent state $|\alpha'\rangle$ phase shifts of
$2\theta$, $4\theta$, $-\theta$, $-3\theta$, $3\theta$, $5\theta$,
$-2\theta$ and $-4\theta$, respectively. After performing the $X $
homodyne detection on $|\alpha'\rangle$, we retain those states with
phase shift $\theta$. Then, photons $b$ and $d$ should pass through
in one pair of the four paths, i.e. $b_{11}$-$d_{11}$,
$b_{12}$-$d_{12}$, $b_{21}$-$d_{21}$, and $b_{22}$-$d_{22}$. At
last, let those two paths originated from the same path via OD link
to an optical multiplexer(OM), which is other part of WDM and can
merge them into one path. For instance, paths $b_{11}$-$b_{12}$,
which both come from $b_{1}$ by OD, are merged into $b'_{1}$ after
passing through OM. After taking above procedures, photons $b$ and
$d$ then lie in the state of the following entangled form:
\begin{eqnarray}
(|\omega_{1}\omega_{1}\rangle+|\omega_{2}\omega_{2}
\rangle)\bigotimes(|b'_{1}d'_{1}\rangle+|b'_{2}d'_{2}\rangle)\;
.\label{shi7}
\end{eqnarray}

\begin{figure}[h]
\captionstyle{center}
\centering
\includegraphics[width=10cm]{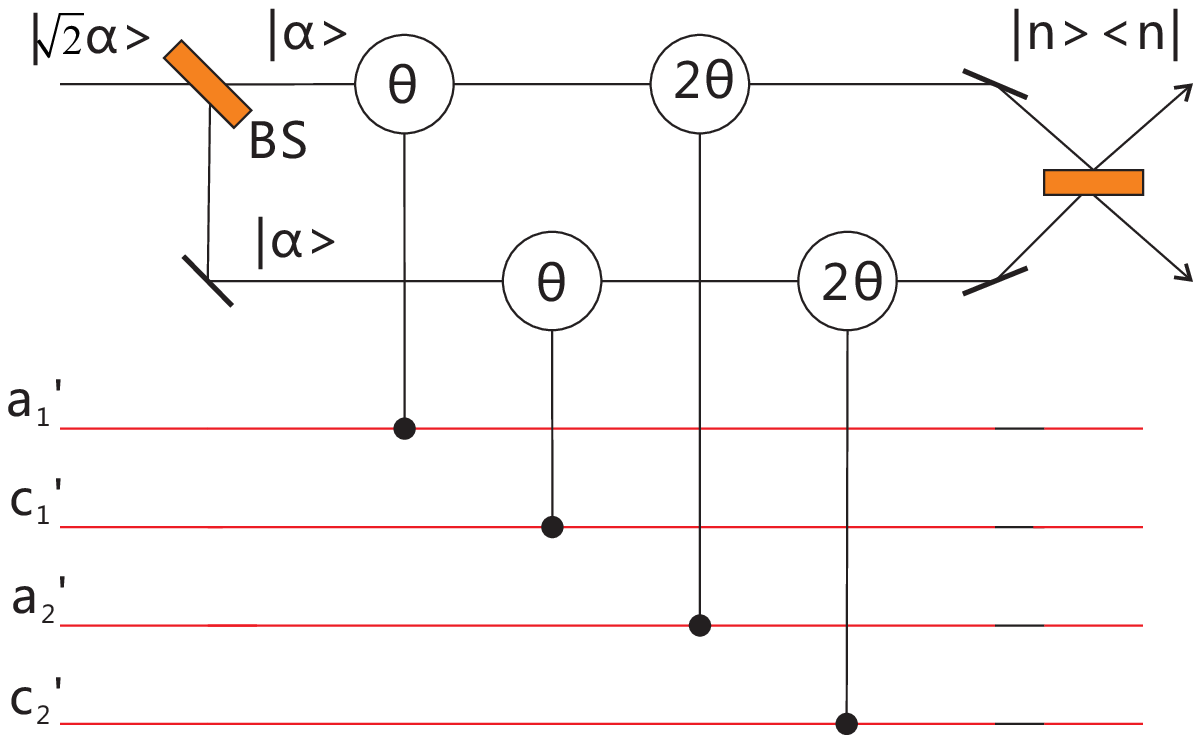}
\caption{Double XPM method to realize the function of
Fig. \ref{fig2}.} \label{fig4}
\end{figure}

\begin{figure}[h]
\captionstyle{center}
\centering
\includegraphics[width=12cm]{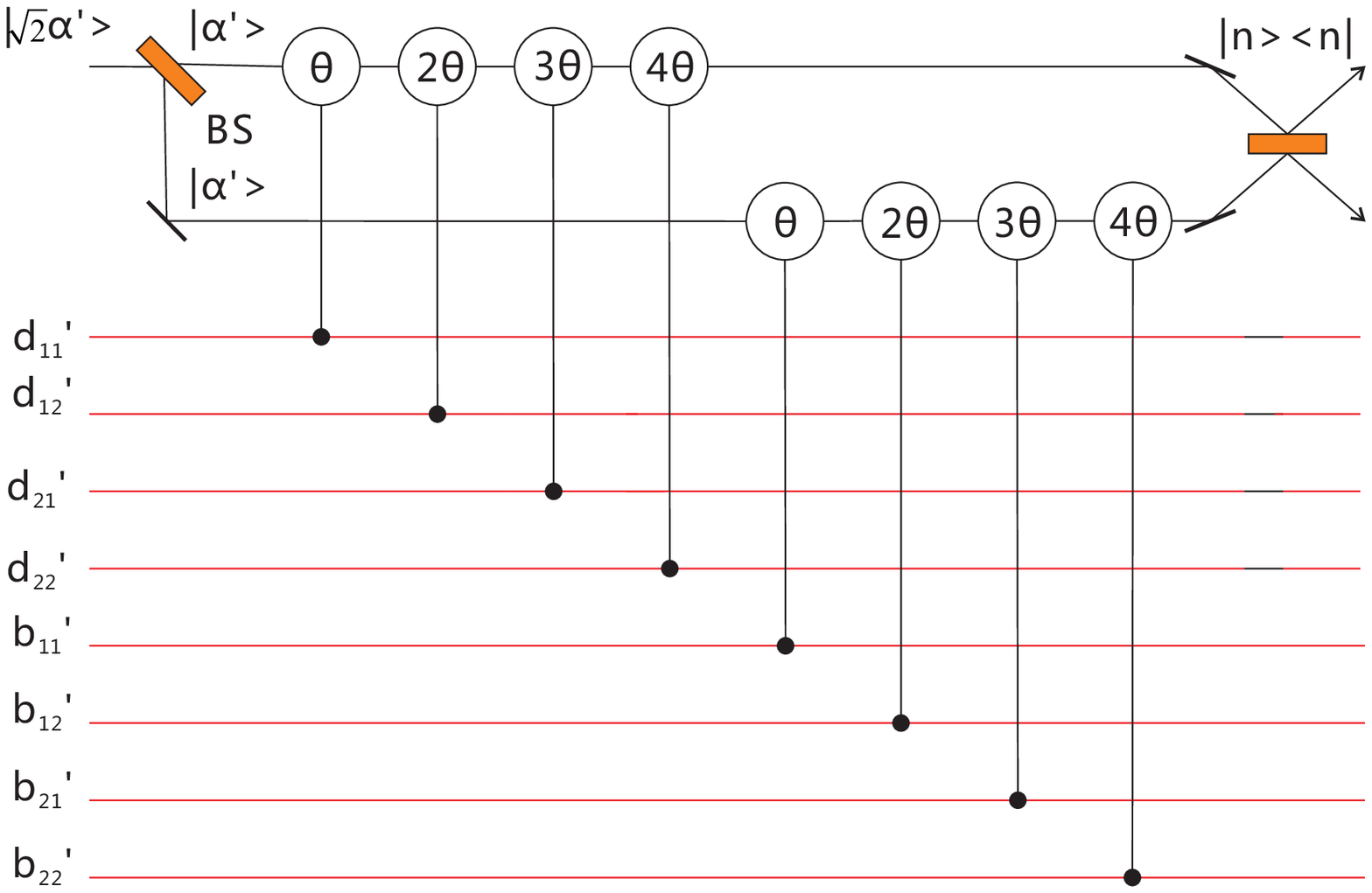}
\caption{Double XPM method to realize the function of
Fig. \ref{fig3}.} \label{fig5}
\end{figure}

In all, from (\ref{shi2}), (\ref{shi3}), (\ref{shi5}) and
(\ref{shi7}), we find that a four-photon 12-qubit hyperentangled
state (\ref{shi1}) is created. Unfortunately, the setups shown in
Figures \ref{fig2} and \ref{fig3} are generally impossible to change
the sign of the phase shift in single nonlinear medium. The signs of
the phase shifts must be identical, since they are all proportional
to the nonlinear susceptibility $\chi^{3}$ \cite{minus phase}. To
circumvent this difficulty, at least one can use either the
cross-kerr with two materials of opposite nonlinearities instead of
the single material, or use double XPM method \cite{double XPM}
without minus phase as shown in Figures \ref{fig4} and \ref{fig5}.
In Figure \ref{fig4}, we use a 50:50 beam splitter (BS) to divide
the coherent state into two beams $|\alpha\rangle$ $|\alpha\rangle$,
and then they are coupled to the photonic modes $a'_{1}$ and
$a'_{2}$, $c'_{1}$ and $c'_{2}$ through the XPM respectively.
Correspondingly, the phase shifts induced by the couplings are
$\theta$ and $2\theta$ in both beams. Afterwards, the two coherent
states after the interaction are compared with a 50-50 BS. Project
the $|n\rangle$$\langle n|$ onto the upper beam, in case $n=0$, the
state (\ref{shi5}) is obtained. Analogously, we can obtain the state
(\ref{shi7}) with the same procedure as presented in Figure
\ref{fig5}.

\begin{figure}[t]
\centering
\includegraphics[width=8cm]{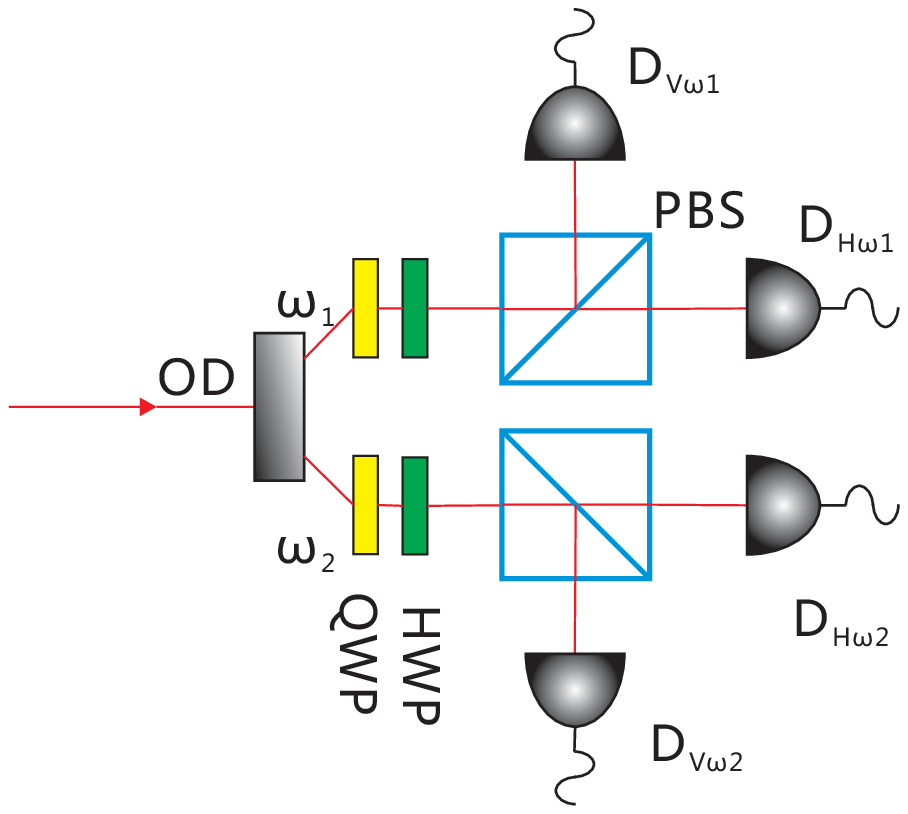}
\caption{Measurement setup. OD: optical demultiplexer; QWP:
quarter-wave plate; HWP: half-wave plate; PBS: polarizing beam
splitter; D: single-photon detector.} \label{fig6}
\end{figure}

A setup for measuring the hyperentanglement in polarization,
frequency and spatial mode simultaneously and independently is
schematically shown in Figure \ref{fig6}. Place eight of these
setups in all eight output paths, the qubits in spatial mode can
then be determined. Of each spatial mode, the optical demultiplexer
splits it into two paths according to frequency, which means one
therefore obtains the qubits in frequency. Subsequently, at both of
these two paths the conventional polarization analysis
\cite{polarization analysis} is implemented, by which the qubits in
polarization are read out.

\section{Quantum characteristics of the hyperentangled state}

To characterize the hyperentangled state (\ref{shi1}), we evaluate
the Bell-type inequality violation of it. Note that the 12-qubit
state (\ref{shi1}) is the eigenstate of $2^{12}$ stabilizing
operators with unit eigenvalue \cite{stabilizing operator,stabilizer
formalism}. i.e.,
\begin{eqnarray}
S_{i}|\psi\rangle=|\psi\rangle, i=1,2,\cdots2^{12}\; ,\label{shi8}
\end{eqnarray}
where \begin{eqnarray}
S_{i}=\bigotimes_{k=1}^{3}O_{j}^{(k)}\; \label{shi9}
\end{eqnarray} describe perfect correlations in the state.
Here, $O_{j}^{(k)}$$\in$\{$I$, $\pm$$X_{j}^{(k)}$,
$\pm$$Y_{j}^{(k)}$, $\pm$$Z_{j}^{(k)}$\} represent the observables
for each photon. $I$ and $X_{j}^{(k)}$, $Y_{j}^{(k)}$, $Z_{j}^{(k)}$
denote the identity and Pauli matrices, with subscript $j=1, 2, 3,
4$ signifying photons $a, b, c$ and $d$, and superscript $k=1, 2, 3$
signifying the three DOF. The operators $X$, $Y$, and $Z$ satisfy
relationship $ZX=-XZ=iY$.

According to Ref.\cite{bell operator} the Bell operator can be
expressed as
\begin{eqnarray}
B=\sum_{i}^{2^{12}}S_{i}\; .\label{shi10}
\end{eqnarray}
The local hidden variable theory(LHVT) sets the upper bound for the
expectation value of $B$, and hence satisfies the following
inequality \cite{Two Observers}
\begin{eqnarray}
\langle B\rangle\leq max_{LHVT}|\langle B\rangle|\; .\label{shi11}
\end{eqnarray}
We notice that among all 4096 $S_{i}$, there are 1854 of them with
negative signs in front as shown in the Appendix, like
\begin{eqnarray*}
-Y_{1}^{(1)}Y_{2}^{(1)}X_{3}^{(1)}X_{4}^{(1)}|\psi\rangle=|\psi\rangle\; .
\end{eqnarray*}

In LHVT, the observables are elements of reality, that means their
values are precisely defined, i.e. either $+1$ or $-1$ \cite{EPR}.
Therefore, the upper bound of the expectation value of $B$ in
(\ref{shi10}) is $2^{12}-2\times1854=388$. On the other hand, in
quantum mechanics, the expectation value of $B$ is $2^{12}=4096$.
Obviously, the QM result greatly violates the inequality
(\ref{shi11}), the bound set by the LHVT.

Following, we compare the degrees of violation of different GHZ-type
states with same amount of qubits but different numbers of photons.
According to the Appendix, one may conclude that the more the
entangled degrees of freedom are, the stronger the violation of
Bell-type inequality will be. To give a concrete example, in Table
\ref{table1} we compare the violation degrees to the Bell-type
inequality of three different 12-qubit entangled states, that is
(\ref{shi1}),
\begin{eqnarray}
|\psi'\rangle=(|HHHHHH\rangle+|VVVVVV\rangle)\bigotimes
(|\omega_{1}\omega_{2}\omega_{1}\omega_{2}\omega_{1}
\omega_{2}\rangle+|\omega_{2}\omega_{1}\omega_{2}\omega_{1}
\omega_{2}\omega_{1}\rangle)\; , \label{shi-2}
\end{eqnarray}
and
\begin{eqnarray}
|\psi''\rangle=(|HHHHHHHHHHHH\rangle+|VVVVVVVVVVVV\rangle)\; .
\label{shi-3}
\end{eqnarray}
The results in the Table indicate that the state (\ref{shi1})
entangled in three DOFs violates the Bell inequality (\ref{shi11})
most, the state (\ref{shi-2}) entangled in two DOFs goes next, and
the state (\ref{shi-3}) entangled in only one DOF violates the Bell
inequality least.

\section{Conclusions}

In summary, in this work we propose a novel scheme for the
generation of simultaneous entanglement of four photons in three
independent DOFs, i.e., a 12-qubit hyperentangled state, by virtue
of the linear optical instrument and cross-Kerr nonlinearity. We
show that the proposed scheme is within the reach of nowadays
technology. Theoretically, our method can be simply expanded to the
production of hyperentangled state with more photons, although
practically the difficulties in the manipulation of six or more
photons make the expansion non-realistic. To check the nonlocality
nature of this 12-qubit hyperentangled state, we evaluate its
violation degree to a Bell-type inequality. The result agrees well
with the QM prediction while extremely contradicts to the LR
constraint, which indicates that the genuine multiparticle
hyperentangled state possesses certain superiority in comparison
with the normal entangled state in the verification of QM
nonlocality, quantum computation, and quantum information
processing.

\vskip 0.7cm
\noindent {\bf Acknowledgments}

This work was supported in part by the National Natural Science
Foundation of China(NSFC), by the CAS Key Projects KJCX2-yw-N29 and
H92A0200S2. We thank Junli Li, Hongbo Xu and Chao Niu for useful discussions.


\pagebreak

\noindent
\appendix{\noindent\bf{Appendix: Comparation of Local Realism Violation}}
\label{app:def}

\noindent Here, we compare the violation degrees to the Bell
inequality of the GHZ-type states:
\begin{eqnarray*}
|\psi_{1}\rangle=(|H_{1}H_{2}\cdots H_{m}\rangle+|V_{1}V_{2}\cdots
V_{m}\rangle)^{\bigotimes{3}}\; ,
\end{eqnarray*}
\begin{eqnarray*}
|\psi_{2}\rangle=(|H_{1}H_{2}\cdots H_{n}\rangle+|V_{1}V_{2}\cdots
V_{n}\rangle)^{\bigotimes{2}}\; ,
\end{eqnarray*}
\begin{eqnarray*}
|\psi_{3}\rangle=(|H_{1}H_{2}\cdots H_{l}\rangle + |V_{1}V_{2}\cdots
V_{l}\rangle)\; ,
\end{eqnarray*}
where the superscripts $3$, $2$, $1$ represent the number of
entangled DOFs; $m$, $n$, $l$ signify the photon numbers, and
satisfy $m\times3$=$n\times2$=$l\times1$, i.e they have the same
qubits. According to the method in calculating the state
(\ref{shi1}) in the text, we just need to calculate the number of
$S_{i}$ with negative sign, labeled as $C(\psi)$. The larger
$C(\psi)$ means a bigger violation.

First, we compute the $C(\psi)$ of the state which has $m$ photons
in one DOF by taking the advantage of equation:
$$C(\psi)=C_{m}^{2}+C_{m}^{6}+\ldots +
C_{m}^{2+4i}\; \eqno{(A1)}$$ with $\ 2i+4\leqslant m$. For different
values of $m$, the above equation tells
$$C(\psi)=\begin{cases}2^{m-2}&\text{if }m=4x-2\\
(-1)^{\frac{m+5}{4}}2^{\frac{m-3}{2}}+2^{m-2}&\text{if }m=4x-1\\
(-1)^{\frac{m+4}{4}}2^{\frac{m-2}{2}}+2^{m-2}&\text{if }m=4x\\
(-1)^{\frac{m+3}{4}}2^{\frac{m-3}{2}}+ 2^{m-2}&\text{if
}m=4x+1\end{cases}\ ,\eqno{(A2)}$$ where $x=1,2\cdots$.

Then, we arrive at the results in following ten different cases:
\begin{enumerate}
\item $m$ is odd and $m=4x-1$.
In this case $n=\frac{3m}{2}$ is not integer, $l=3m=12x-3$, so

$C(\psi_{1})-C(\psi_{3})=3((-1)^{\frac{x+1}{4}}2^{2x-2}+
2^{4x-3})(2^{4x-1}-(-1)^{\frac{x+1}{4}}2^{2x-2}-2^{4x-3})^{2}+
((-1)^{\frac{x+1}{4}}2^{2x-2}+2^{4x-3})^{3}-((-1)^{3x}2^{6x-3}+2^{12x-5})
 =-3\times2^{-7+6x}\times(8\times(-1)^{x}+(-1)^{x}
 \times2^{1+4x}+4^{1+x}-64^{x})>0$, for $x=1,2,\cdots$.

\item $m$ is odd and $m=4x+1$. In this case
$n=\frac{3m}{2}$ is not integer, $l=3m=12x+3$, so

 $C(\psi_{1})-C(\psi_{3})=3((-1)^{\frac{x+1}{4}}
 2^{2x-1}+2^{4x-1})(2^{4x+1}-(-1)^{\frac{x+1}{4}}2^{2x-1}-2^{4x-1})^{2}+
 ((-1)^{\frac{x+1}{4}}2^{2x-1}+2^{4x-1})^{3}-((-1)^{3x+2}2^{6x}+2^{12x+1})
=-3\times2^{-1+6x}\times((-16)^{x}+(-1)^{x}+
4^{x}-64^{x})>0$ for $x=1,2,\cdots$.

\item $m$ is even and $m=8x+2$. In this case $n=\frac{3m}{2}=12x+3$, so

$C(\psi_{1})-C(\psi_{2})=3(2^{8x})(2^{8x+2}-2^{8x})^{2} +
(2^{8x})^{3}-2((-1)^{3x+2}2^{6x}+2^{12x+1})(2^{12x+3}-
(-1)^{3x+2}2^{6x}-2^{12x + 1}) = 2^{1+12x} +
4^{1+12x}-(-1)^{x}\times8^{1+6x}>0$ for $x=0,1,\cdots$.

\item $m$ is even and $m=8x-2$. In this case $n=\frac{3m}{2}=12x-3$, so

$C(\psi_{1}) - C(\psi_{2})=3(2^{8x-4})(2^{8x-2}-2^{8x-4})^{2} +
(2^{8x-4})^{3}-2((-1)^{3x}2^{6x-3}+2^{12x-5})
(2^{12x-3}-(-1)^{3x}2^{6x-3}-2^{12x-5}) =
4^{-5+6x}\times(32+4^{6x}-(-1)^{x}4^{2+3x})>0$ for $x=1,2,\cdots$.

\item $m$ is even and $m=8x+4$. In this case $n=\frac{3m}{2}=12x+6$, so

$C(\psi_{1})-C(\psi_{2})=3((-1)^{2x+2}2^{4x+1}+2^{8x+2})(2^{8x+4}-
(-1)^{2x+2}2^{4x+1}-2^{8x+2})^{2}+
((-1)^{2x+2}2^{4x+1}+2^{8x+2})^{3}-2(2^{12x+4})(2^{12x+6}-2^{12x+4})
=2^{5+12x}\times(1-3\times2^{1+4x}+3\times4^{1+4x}+8^{1+4x})>0$ for
$x=0,1,\cdots$.

\item $m$ is even and $m=8x$. In this case $n=\frac{3m}{2}=12x$, so

$C(\psi_{1})-C(\psi_{2})=3((-1)^{2x+1}2^{4x-1}+2^{8x-2})(2^{8x}-
(-1)^{2x+1}2^{4x-1}-2^{8x-2})^{2}+
((-1)^{2x+1}2^{4x-1}+2^{8x-2})^{3}-2((-1)^{3x+1}2^{6x-1} +
2^{12x-2})(2^{12x}-(-1)^{3x+1}2^{6x-1}-2^{12x-2})
=16^{-1+4x}\times(-12+(-1)^{x}
\times2^{3+2x}-3\times2^{1+4x}+256^{x})>0$ for $x=1,2,\cdots$.

\item $n$ is even and $n=4x-2$. In this case $l=2n=8x-4$, so

$C(\psi_{2})-C(\psi_{3})=2(2^{4x-4})(2^{4x-2}-2^{4x-4}-
((-1)^{2x}2^{4x-3}+2^{8x-6}) =2^{-7+4x}\times(-16+16^{x})>0$ for
$x=2,3,\cdots$.

\item $n$ is even and $n=4x$. In this case $l=2n=8x$, so

$C(\psi_{2})-C(\psi_{3})=2((-1)^{x+1}2^{2x-1}+2^{4x-2})
(2^{4x}-(-1)^{x+1}2^{2x-1}-2^{4x-2})-((-1)^{2x+1}2^{4x-1}+2^{8x-2})
= 8^{-1+2x}\times(-4\times(-1)^{x}+4^{x})>0$ for $x=1,2,\cdots$.

\item $n$ is odd and $n=4x-1$. In this case $l=2n=8x-2$, so

$C(\psi_{2})-C(\psi_{3})=2((-1)^{x+1}2^{2x-2}+2^{4x-3})
(2^{4x-1}-(-1)^{x+1}2^{2x-2}-2^{4x-3})-(2^{8x-4})
=2^{-5+4x}\times(-4-(-1)^{x}\times4^{1+x}+16^{x})>0$ for
$x=1,2,\cdots$.

\item $n$ is odd and $n=4x+1$. In this case $l=2n=8x+2$, so

$C(\psi_{2})-C(\psi_{3})=2((-1)^{x+1}2^{2x-1}+2^{4x-1})
(2^{4x+11}-(-1)^{x+1}2^{2x-1}-2^{4x-1})-(2^{8x}) =
2^{-1+4x}\times(-1-(-1)^{x}\times2^{1+2x}+16^{x})>0$ for
$x=1,2,\cdots$.
\end{enumerate}

In conclusion, it is evident that $C(\psi_{1})
> C(\psi_{2}) > C(\psi_{3})$, which tells that for the three GHZ-type
entangled states with same qubits, the one entangled in three DOFs,
the $\psi_1$, violates LR most, the state $\psi_2$ entangled in two
DOFs violates the second, and the state $\psi_3$ entangled only in
one DOF violates the least. \label{lastpage}


\newpage

\begin{table}[t]
\captionstyle{flushleft}
\caption{Comparation of violation degree for different entangled
states.}
\begin{tabular}{ccccc}
\hline\hline \hspace{.9cm}state\hspace{.9cm} &\hspace{.6cm}$\langle
B\rangle_{QM}$ \hspace{.6cm}&\hspace{.9cm}$max_{LHVT}|\langle
B\rangle$\hspace{.9cm} & \hspace{.9cm}$D=\langle B
\rangle_{QM}/max_{LHVT}|\langle B\rangle$\hspace{.9cm}\\
\hline
$\psi$&       $4096$ &     $388$ &      $10.56$&\\
$\psi'$&       $4096$ &     $1024$ &      $4.00$&\\
$\psi''$&       $4096$ &     $1984$ &      $2.06$&\\
\hline\hline
\end{tabular}
\label{table1}
\end{table}

\newpage

\newpage
\begin{thebibliography}{99}

\bibitem{quantum-computation} Nielsen, M.A. and Chuang, I.L., \
{\it Quantum Computation and Quantum Information}\ (Cambridge
University Press, Cambridge, England, {\bf 2000}).

\bibitem{one-way quantum computer} Raussendorf, R.; Briegel,
H. J., {\it Phys. Rev. Lett.} {\bf 2001,} 86, 5188-5191.

\bibitem{C. Wang et al} Wang, C.; Deng, F.G.;Li, Y.S.; Liu,
X.S. and Long, G.L., {\it Phys. Rev. A} {\bf 2005,} 71, 044305.

\bibitem{Teleporting} Bennett, C.H.; Brassard, C.; Cr\'{e}peau,
C.; Jozsa, R.; Peres, A.; Wootters, W.K., {\it Phys. Rev. Lett.}
{\bf 1993,} 70, 1895.

\bibitem{Purification using Spatial Entanglement} Simon,
C.; Pan, J.W., {\it Phys. Rev. Lett.} {\bf 2002,} 89, 257901.

\bibitem{D. Bruss and C. Macchiavello} Bruss,
D.; Macchiavello, C., {\it Phys. Rev. Lett.} {\bf 2002,} 88, 127901.

\bibitem{nature Pan} Gao, W.B.; Lu, C.Y.; Yao, X.C.; Xu,
P.; G\"{u}hne, O.; Goebel, A.; Chen, Y.A.; Peng, C.Z.; Chen, Z.B.;
Pan, J.W., {\it Nature Physics} {\bf 2010,} 6, 331.

\bibitem{Kwiat P. G} Kwiat, P. G., {\it J. Mod. Opt.} {\bf 1997,} 44, 2173.

\bibitem{Ultrabright source} Kwiat, P.G.; Waks, E.; White,
A.G.; Appelbaum, I.; Eberhard, P.H., {\it Phys. Rev. A} {\bf 1999,}
60, R773.

\bibitem{Polarization-momentum} Barbieri, M.; Cinelli, C.; Mataloni,
P.; De-Martini, F., {\it Phys. Rev. A} {\bf 2005,} 72, 052110.

\bibitem{two-photon four-qubit} Chen, K.; Li, C.M.; Zhang,
Q.; Chen, Y.A.; Goebel, A.; Chen, S.; Mair, A.; Pan, J.W., {\it
Phys. Rev. Lett.} {\bf 2007,} 99, 120503.

\bibitem{All-Versus-Nothing} Yang, T.; Zhang, Q.; Zhang, J.; Yin,
J.; Zhao, Z.; \'{Z}ukowski, M.; Chen, Z.B.; Pan, J.W., {\it Phys.
Rev. Lett.} {\bf 2005,} 95, 240406.

\bibitem{All-Versus-Nothing2} Chen, Z.B.; Pan, J.W.; Zhang, Y.D.;
\v{C}aslav Brukner; Zeilinger, A., {\it Phys. Rev. Lett.} {\bf
2003,} 90, 160408.

\bibitem{two-photon six-qubit} Vallone, G.; Ceccarelli, R.; Martini,
F.D.; Mataloni, P., {\it Phys. Rev. A} {\bf 2009,} 79, R030301.

\bibitem{Hyperentangled Photon Pairs} Barreiro, J.T.; Langford, N.K.; Peters,
N.A.; Kwiat, P.G., {\it Phys. Rev. Lett.} {\bf 2005,} 95, 260501.

\bibitem{Schuck-96-2006} Schuck, C.; Huber, G.; Kurtsiefer, C. and Weinfurter,
H., {\it Phys. Rev. Lett.} {\bf 2006,} 96, 190501.

\bibitem{Barreiro-4-2008} Barreiro, J. T.; Wei, T. C. and Kwiat, P. G.. {\it Nature
Phys.} {\bf 2008,} 4, 282.

\bibitem{Vallone-100-2008} Vallone, G.; Pomarico, E.; Martini, F. De and Mataloni, P.,
{\it Phys. Rev. Lett.} {\bf 2008,} 100, 160502.

\bibitem{Briegel-86-2001} Briegel, H. J. and Raussendorf, R., {\it Phys. Rev. Lett.}
{\bf 2001,} 86, 910.

\bibitem{Barbieri-97-2006} Barbieri, M.; Martini, F. D.; Mataloni, P.; Vallone, G.
and Cabello, A., {\it Phys. Rev. Lett.} {\bf 2006,} 97, 140407.

\bibitem{ghz1990} Greenberger, D.M.; Horne, M.; Shimony, A.; Zeilinger,
A., {\it Am. J. Phys.} {\bf 1990,} 58, 1131.

\bibitem{BBO} Kwiat, P.G.; Mattle, K.; Weinfurter, H.; Zeilinger,
A., {\it Phys. Rev. Lett.} {\bf 1995,} 75, 4337.

\bibitem{SPDC} Walborn, S.P.; De-Oliveira, A.N.; Thebaldi, R.S.; Monken,
C.H., {\it Phys. Rev. A} {\bf 2004,} 69, 023811.

\bibitem{frequency-entangled} Yabushita, A.; Kobayashi, T.,
{\it Phys. Rev. A} {\bf 2004,} 69, 013806.

\bibitem{cross-kerr} Sheng, Y.B.; Deng, F.G.; Zhou, H.Y.,
{\it Phys. Rev. A} {\bf 2008,} 77, 042308.

\bibitem{quantum communication} Sheng, Y.B.; Deng, F.G.; Long, G.L.,
{\it Phys. Rev. A} {\bf 2010,} 82, 032318.

\bibitem{homodyne} Nemoto, K.; Munro, W.J.,
{\it Phys. Rev. Lett.} {\bf 2004,} 93, 250502.

\bibitem{WDM} Sheng, Y.B.; Deng, F.G.,
{\it Phys. Rev. A} {\bf 2010,} 81, 032307.

\bibitem{minus phase} Kok, P.,
{\it Phys. Rev. A} {\bf 2008,} 77, 013808.

\bibitem{double XPM} He, B.; Ren, Y.H.; Bergou, J.A.,
{\it Phys. Rev. A} {\bf 2009,} 79, 052323.

\bibitem{polarization analysis} Vallone, G.; Pomarico, E.;
Mataloni, P.; De-Martini, F.; Berardi, V., {\it Phys. Rev. Lett.}
{\bf 2007,} 98, 180502.

\bibitem{stabilizing operator} Cabello, A.; Rodriguez, D.; Villanueva,
I., {\it Phys. Rev. Lett.} {\bf 2008,} 101, 120402.

\bibitem{stabilizer formalism} G\'{e}za-T\'{o}th; G\"{u}hne, O.,
{\it Phys. Rev. A} {\bf 2005,} 72, 022340.

\bibitem{bell operator}G\"{u}hne, O.; G\'{e}za-T\'{o}th; Hyllus,
P.; Briegel, H.J., {\it Phys. Rev. Lett.} {\bf 2005,} 95, 120405.

\bibitem{Two Observers}Cabello, A., {\it Phys. Rev. Lett.}
{\bf 2001,} 87, 010403.

\bibitem{EPR}Einstein, A.; Podolsky, B.; Rosen, N., {\it Phys. Rev.}
{\bf 1935,} 47, 777.

\end{thebibliography}
\end{document}